\documentstyle[12pt]{article}
\makeatletter

\if@twoside
Probably too small to be useful
\else
\fi
\ifcase \@ptsize
\or 
\or 
\fi

\def\section{\@startsection {section}{1}{\z@}{-3.5ex plus -1ex minus
 -.2ex}{2.3ex plus .2ex}{\bf\raggedright}}
\def\subsection{\@startsection{subsection}{2}{\z@}{-3.25ex plus -1ex
minus
 -.2ex}{1.5ex plus .2ex}{\sc\raggedright}}
\def\bibliography#1{\if@filesw\immediate\write\@auxout
  {\string\bibstyle{npb}}\fi
  \if@filesw\immediate\write\@auxout{\string\bibdata{#1}}\fi
  \@input{\jobname.bbl}}
\def\thebibliography#1{
  \section{References\@mkboth{REFERENCES}{REFERENCES}}\list
{[\arabic{enumi}]}{\settowidth\labelwidth{[#1]}\leftmargin\labelwidth
 \advance\leftmargin\labelsep \itemsep=0pt
 \usecounter{enumi}}
 \def\newblock{\hskip .11em plus .33em minus .07em}
 \sloppy\clubpenalty4000\widowpenalty4000
 \sfcode`\.=1000\relax}
\def\@citex[#1]#2{%
\if@filesw \immediate \write \@auxout {\string \citation {#2}}\fi
\@tempcntb\m@ne \let\@h@ld\relax \def\@citea{}%
\@cite{%
  \@for \@citeb:=#2\do {%
    \@ifundefined {b@\@citeb}%
      {\@h@ld\@citea\@tempcntb\m@ne{\bf ?}%
      \@warning {Citation `\@citeb ' on page \thepage \space
undefined}}%
      {\@tempcnta\@tempcntb \advance\@tempcnta\@ne%
      \@tempcntb\number\csname b@\@citeb \endcsname \relax%
      \ifnum\@tempcnta=\@tempcntb 
        \ifx\@h@ld\relax%
          \edef \@h@ld{\@citea\csname b@\@citeb\endcsname}%
        \else%
          \edef\@h@ld{\ifmmode{-}\else--\fi\csname
b@\@citeb\endcsname}%
        \fi%
      \else
        \@h@ld\@citea\csname b@\@citeb \endcsname%
        \let\@h@ld\relax%
      \fi}%
    \def\@citea{,\penalty\@highpenalty\,}%
  }\@h@ld%
}{#1}}
\newif\ifabstract \abstractfalse
\newif\ifconsolidatetitle \consolidatetitlefalse
\def\ds@consolidatetitle{\consolidatetitletrue}
\gdef\@publabel{\hfil}
\gdef\@pubdate{\null}
\gdef\@pubnumber{\null}
\gdef\@author{\null}
\gdef\@title{\null}
\gdef\@abstract{\null}
\long\def\pubdate#1{\gdef\@pubdate{#1}}
\long\def\pubnumber#1{\gdef\@pubnumber{#1}}
\long\def\publabel#1{\gdef\@publabel{#1}}
\long\def\author#1{\gdef\@author{#1}}
\long\def\title#1{\gdef\@title{#1}}
\long\def\abstract#1{\abstracttrue\gdef\@abstract{#1}}
\def\titlerelax{
\let\maketitle\relax
\let\settitleparameters\relax
\let\consolidatetitle\relax
\let\inittitlepage\relax
\let\finishtitlepage\relax
\let\titlepagecontents\relax
\let\multithanks\relax
\let\titlebaselines\relax
\let\@makepub\relax
\let\@maketitle\relax
\let\@makeauthor\relax
\let\@makeabstract\relax
\let\@maketitlenote\relax
\let\thanks\relax
\let\titlerelax\relax}
\def\titleclean
{\gdef\@titlenote{}
\gdef\@abstract{}
\gdef\@author{}
\gdef\@title{}
\gdef\@pubdate{}\gdef\@pubnumber{}\gdef\@publabel{}
\gdef\@dpublabel{}
}

\def\@makepub{\vbox to \z@{\hbox to \textwidth{\hfill
\@publabel \hfill
\llap{\parbox[t]{0.25\textwidth}{\raggedleft\@pubnumber}}}%
\vss}}
\def\@maketitle{\vskip 60pt \begin{center}
 {\LARGE \@title \par}
 \end{center}}
\def\@makeauthor{{%
\def\and{\smallskip {\normalsize \rm and\smallskip }}
\def\And{\medskip {\normalsize \rm and\\}\medskip}
\long\def\address##1{{\def\and{\\and\\}\medskip
				{\small \it \\##1\\}
}}
{\centering
 \vskip 3em
 \large \lineskip .75em
 \@author}
 \par}}
\def\@makedate{\vskip 1.5em
 {\raggedright \small \noindent\@pubdate \par}}
\def\@makeabstract{\vskip 1.5em
{\small
\begin{center}
{\bf ABSTRACT\vspace{-.5em}\vspace{0pt}}
\end{center}
\quotation \@abstract \endquotation}}
\def\@consolidatetitle{{
\thispagestyle{empty}
\@makepub
\setlength{\parskip}{0pt}
\null
\vskip 7mm
\nointerlineskip
\@maketitle
\vskip 5ex
\@makeauthor
\vskip 4 ex
\@makeabstract
\vskip 5 ex
}}

\def\maketitle{\ifconsolidatetitle\@consolidatetitle
\else
	\titlepage
	\let\footnotesize\small \setcounter{page}{0}
	\@makepub
	\vfil
	\@maketitle
	\@makeauthor
	\vfil
	\ifabstract\@makeabstract\fi
	\@thanks
	\vfil
	\@makedate
	\if@restonecol\twocolumn\fi \eject
\fi
	\titlerelax \titleclean
	\setcounter{footnote}{0}
}

\def\square{\hbox{\vrule height2ex\kern-0.4pt
\vbox to 2ex{\hrule width2ex\vfil\hrule width2ex}\kern-0.4pt\vrule
height2ex}}

{\unskip\nobreak\hfill$\Box$\end{list}}

\def\lefteqn#1{\hbox to 0pt{$\displaystyle #1$\hss}}
\def\starteqn#1#2{\hbox to 0pt{${\displaystyle #1}\textstyle #2$\hss}
\qquad\nonumber\\*&#2&}
\def\blank#1{}

\def\en{\end{equation}}

\def\enn{\end{eqnarray}}

\def\eq{\begin{equation}}
\def\eqq{\begin{eqnarray}}

\def\eqq{\begin{eqnarray}}
\def\enn{\end{eqnarray}}

\ifcase\@ptsize
 \font\tenmsa=msam10
 \font\sevenmsa=msam7
 \font\fivemsa=msam5
 \font\tenmsb=msbm10
 \font\sevenmsb=msbm7
 \font\fivemsb=msbm5
\or
 \font\tenmsa=msam10 scaled \magstephalf
 \font\sevenmsa=msam8
 \font\fivemsa=msam6
 \font\tenmsb=msbm10 scaled \magstephalf
 \font\sevenmsb=msbm8
 \font\fivemsb=msbm6
\or
 \font\tenmsa=msam10 scaled \magstep1
 \font\sevenmsa=msam8
 \font\fivemsa=msam6
 \font\tenmsb=msbm10 scaled \magstep1
 \font\sevenmsb=msbm8
 \font\fivemsb=msbm6
\fi

\newfam\msafam
\newfam\msbfam
\textfont\msafam=\tenmsa  \scriptfont\msafam=\sevenmsa
  \scriptscriptfont\msafam=\fivemsa
\textfont\msbfam=\tenmsb  \scriptfont\msbfam=\sevenmsb
  \scriptscriptfont\msbfam=\fivemsb

\def\hexnumber@#1{\ifnum#1<10 \number#1\else
 \ifnum#1=10 A\else\ifnum#1=11 B\else\ifnum#1=12 C\else
 \ifnum#1=13 D\else\ifnum#1=14 E\else\ifnum#1=15
F\fi\fi\fi\fi\fi\fi\fi}

\def\msa@{\hexnumber@\msafam}
\def\msb@{\hexnumber@\msbfam}
\def\Bbb{\ifmmode\let\next\Bbb@\else
 \def\next{\errmessage{Use \string\Bbb\space only in math
mode}}\fi\next}
\def\Bbb@#1{{\Bbb@@{#1}}}
\def\Bbb@@#1{\fam\msbfam#1}

\makeatother

\begin{document}

%









\pubnumber{ cond-mat/yymmnn }

\pubdate{July 4, 1992}

\title{On the Harmonic Oscillator Regularization\\ of Partition
Functions}

\author{
       K{\aa}re~Olaussen$^{1}$
\address{
       Gruppe for Teoretisk Fysikk,\\
       Universitetet i Trondheim, NTH,\\
       N-7034 Trondheim, NORWAY}
}

\footnotetext[1]{
Email: {\tt kolausen@phys.unit.no}
}

\abstract{
A convenient way to calculate $N$-particle quantum partition
functions is
by confining the particles in a weak harmonic potential instead of
using a
finite box or periodic boundary conditions. There is, however, a
slightly
different connection between partition functions and thermodynamic
quantities with such volume regularization. This is made explicit,
and
its origin explained to be due to the system having a space-varying
density
in an external potential.
Beyond perturbation theory there is a potential pitfall with the
method, which
is pointed out.
\\
\
\\
\
}

\maketitle

In order to simplify the computation of quantum partition functions
one
would prefer to utilize a geometry which allows for separation of
center-of-mass motion and preserves as many symmetries of the model
as
possible. Using a finite box to confine particles does not meet the
first
requirement, and periodic boundary conditions are in conflict with
the usual
rotiational invariance of particle interactions. A rather convenient
solution
is to confine the particles in a harmonic potential, and
mimick the infinite volume limit by letting this potential vanish. By
comparing
the partition function for a single particle in a finite box of
($D$-dimensional)
volume $V$,
\[
    Z_1^{(V)} = {V}\,{\Lambda^{-D}_{T}} =
V\,\left({m}/{2\pi\beta\hbar^2} \right)^{D/2},
    \ \ \ \mbox{with $\beta=1/k_B T$,}
\]
and in the harmonic oscillator potential $\Phi(r)=\frac{1}{2} m
\omega^2 r^2$,
\[
    Z_1^{({\omega})} = \left[{2
\sinh(\frac{1}{2}\beta\hbar\omega)}\right]^{-D} \approx
    \left({\beta\hbar \omega}\right)^{-D},
\]
one finds a volume $V_{\omega}$ corresponding to a given potential
strenght $m \omega^2$:
\begin{equation}
    V_{\omega} = \left({2\pi}/{\beta m \omega^2}\right)^{D/2}.
                                                 \label{Veff}
\end{equation}
$V_{\omega}$ can be used as an effective volume in the calculation of
e.g.\ virial coefficients,
as was first done by Comtet {\em et.\ al.}\cite{Comtet} for the
second virial coefficient of the anyon{\cite{Leinaas,Wilczek}} gas.
Their result is in agreement with an earlier calculation by Arovas
{\em et.\ al.}\cite{Arovas} (who
used confinement in a circular disk), but the analysis is much
simplified.
The same method has been applied by Blum {\em et.\ al.}\cite{Blum} to
a similar anyon model,
and is also used in many other analyses of anyon
systems\cite{Sporre,Sen}.
But there is nothing in the method which restricts its applicability
to such
systems only,---one expects it to be a convenient method for volume
regularization in general.

The connection between the thermodynamic quantities and partition
functions is however
different in these two cases. This fact, and its origin, does not
seem to have been explicitly
commented on in the literature.
To review briefly, all intrinsic thermodynamic quantities can in
principle be derived from
the fugacity expansion of the pressure
\begin{equation}
      \beta {p} = \sum_{\ell=1}^{\infty} \, b_{\ell} \, z^{\ell},
                                     \label{pressure}
\end{equation}
where $z=\exp(\beta\mu)$, with $\mu$ the chemical potential.
For instance, the density is given as
\begin{equation}
      \rho = \beta z {\partial p}/{\partial z} =
      \sum_{\ell=1}^{\infty} \, \ell\, b_{\ell} \; z^{\ell},
                                    \label{density}
\end{equation}
{}from which the virial expansion can be obtained (by inverting
(\ref{density}) into
an expansion for $z=z(\rho)$, and inserting the latter into
(\ref{pressure})). I.e.
\[
   \beta p = \sum_{\ell=1}^{\infty} \; a_\ell\,\rho^{\ell},
\]
with
\[
   a_1 = \frac{1}{b_1},\ \
   a_2 = -\frac{b_2}{b_1^2},\ \
   a_3= \frac{4 b_2^2-2 b_1 b_3}{b_1^4},\ \
   a_4= -\frac{20 b_2^3 - 18 b_1 b_2 b_3 + 3 b_1^2 b_4}{b_1^6}\ \
\mbox{etc.}
\]
Likewise
the free energy per volume unit $f$ is found from the relation
\[
     \beta f = \rho\,\log\left( z(\rho) \right) - \beta p =
     \rho \log\left( b_1^{-1}\,\rho \right) +
\sum_{\ell=2}^{\infty}\;
     \frac{1}{\ell-1} a_{\ell}\,\rho^{\ell}.
\]
With confinement in a finite box of volume $V$ the connection between
$n$-particle partition functions
$Z_{n}^{(V)}$ and the $b_{\ell}$'s follows from the relation
\begin{equation}
    \exp( \beta p V) = \sum_{n=0}^{\infty} \, z^n \, Z_n^{(V)} \equiv
    \exp\left( \sum_{\ell=1}^{\infty} \, z^{\ell}\, U_{\ell}(
Z_1^{(V)},\ldots,Z_{\ell}^{(V)}\right).
                         \label{KanPart1}
\end{equation}
I.e., by comparing (\ref{KanPart1}) and (\ref{pressure}),
\begin{equation}
    V\,b_{\ell} = U_{\ell}(Z_1^{(V)},\ldots,Z_{\ell}^{(V)}).
                          \label{Mayer1}
\end{equation}
Explicitly, the first few $U_{\ell}$'s are
\begin{eqnarray*}
   &&U_1 = Z_1,\ \ \
     U_2 = \frac{1}{2}\left[2 Z_2 - Z_1^2 \right],\ \ \
     U_3 = \frac{1}{3}\left[3 Z_3 - 3 Z_1 Z_2 + Z_1^3 \right],\\
   &&U_4 = \frac{1}{4}\left[4 Z_4 - 4 Z_1 Z_3 - 2 Z_2^2 + 4 Z_1^2 Z_2
- Z_1^4 \right].
\end{eqnarray*}
However, with confinement in a harmonic potential equation
(\ref{Mayer1}) must be
changed to
\begin{equation}
    V_{\omega}\, \ell^{-D/2}\, b_{\ell} =
U_{\ell}(Z_1^{(\omega)},\ldots,Z_{\ell}^{(\omega)}),
                          \label{Mayer2}
\end{equation}
where $D$ is the number of space dimensions. The reason for the
additional factor
$\ell^{-D/2}$ on the left hand side is that we are dealing
with a system of spatially varying density and pressure. The relation
(\ref{KanPart1})
should thus be modified to
\begin{equation}
    \exp\left( \int d^D r\, \beta p(r) \right) =
    \sum_{n=0}^{\infty}\, z^n\; Z_n^{(\omega)} =
    \exp\left(\sum_{\ell=1}^{\infty} z^{\ell} U_{\ell} \right),
                  \label{KanPart2}
\end{equation}
while the relations between local pressure $p(r)$, density $\rho(r)$,
and fugacity $z$ become
\begin{equation}
    \beta p(r) = \sum_{\ell=1}^{\infty}\; b_{\ell}\,
    \left(z {\rm e}^{-\beta\Phi(r)} \right)^{\ell},\ \ \
    \rho(r) = \sum_{\ell=1}^{\infty}\; \ell\,b_{\ell}\,
    \left(z {\rm e}^{-\beta\Phi(r)} \right)^{\ell}.
                         \label{pressure2}
\end{equation}
These relations are independent of the form of the potential
$\Phi(r)$, provided the latter is
sufficiently well-behaved.
By inserting (\ref{pressure2}) into the left hand side of
(\ref{KanPart2}), with $\Phi(r)=\frac{1}{2}m\omega^2 r^2$,
and performing the space integration one obtains the announced
equation (\ref{Mayer2}),
with $V_{\omega}$ as defined by equation (\ref{Veff}). The assumption
behind this derivation is that the
potential has neglectible variation on the scale of the correlation
length of the system,
so that the local equations of state are equal to those in a
homogeneous bulk system of the same
density.

As a simple consistency check one may compute the partions functions
for ideal Bose and Fermi gases
in the two geometries. The $U_{\ell}$'s are then found to be
\[
     U_{\ell} = \frac{(\pm1)^{\ell+1}}{\ell} \sum_{i} {\rm
e}^{-\ell\beta\epsilon_i}
\]
for ${\mbox{\small bosons}}\atop{\mbox{\small fermions}}$
respectively.
Here the sum runs over all single particle energy
levels, with $\epsilon_i$ the energy of level $i$:
\begin{equation}
  \sum_i {\rm e}^{-\ell\beta\epsilon_i} =
  \left\{
  \begin{array}{ll}
     V \int \frac{d^D k}{(2\pi)^D} \, {\rm e}^{-\ell\beta\hbar^2
k^2/m} =
     V \,\Lambda_T^{-D}\,\ell^{-D/2}&
     \mbox{(periodic boundary conditions)}\\
     & \\
     \left(\sum_n {\rm e}^{-\ell\beta\hbar\omega(n+1/2)}\right)^D
     \approx V_{\omega}\,\Lambda_T^{-D}\,\ell^{-D}
     &\mbox{(harmonic potential)}
  \end{array}
  \right.
\end{equation}
Combined with equations (\ref{Mayer1}) and (\ref{Mayer2})
respectively
one obtains the same expression for $b_\ell = (\pm 1)^{\ell+1}\,
\Lambda_T^{-D}\,\ell^{-(1+D/2)}$ in
the two cases.
However, one notes a difficulty with the method in the case of bosons
below the Bose condensation point. Above Bose condensation the local
equations of
state can be written in the parametric form
\begin{eqnarray*}
   &&\beta p(r) = \left[\Gamma({\textstyle
\frac{D}{2}})\,\Lambda_T^D\right]^{-1}\;
    \int_0^{\infty} d\tau \,\tau^{D/2-1}
    \,\log\left(1 - z\,{\rm e}^{-\beta\Phi(r)}\,{\rm
e}^{-\tau}\right),\\
   &&\rho(r) = \left[\Gamma({\textstyle
\frac{D}{2}})\,\Lambda_T^D\right]^{-1}\;
    \int_0^{\infty} d\tau \,\tau^{D/2-1}
    \, \frac{z}{{\rm e}^{\beta\Phi(r)+\tau} - z }.
\end{eqnarray*}
When $\Phi=0$ at its minimum the fugacity $z$ is restricted to $z \le
1$. This leads to
a maximal value $\rho_{\max}(r)$ for the local density in space
dimensions $D>2$, and this
value is smaller than the bulk critical density where $\Phi(r) > 0$.
To
increase the density further a macroscopic number of particles must
Bose condense into
the ground state. These condensed particles will be spatially
distributed as
\[
    \rho_0(r) = N_0 \, \left(\frac{1}{\pi r_c^2}\right)^{D/2} \, {\rm
e}^{-(r/r_c)^2},
\]
with $N_0$ the number of condensed particles, and
$r_c=\left(\hbar/m\omega\right)^{1/2}$. The ``coherence length''
$r_c$ diverges as $\omega\to 0^+$,
but much slower that the linear system size,
$r_{\omega}=V_{\omega}^{1/D}=(2\pi/\beta m \omega^2)^{1/2}$:
$r_c/r_{\omega}=(2\pi/\beta\hbar\omega)^{1/2}$. In the coherence
region around the
bottom of the potential the local equations of state are different
{}from those in a
homogenous bulk medium of the same density. This may be attributed to
long
range correlations in this region,
\[
   \langle \;\rho_0(r)\, \rho_0(s)\; \rangle -
\langle\;\rho_0(r)\;\rangle \,
   \langle\;\rho_0(s)\;\rangle =
   N_0 \, \left(\frac{1}{\pi r_c^2}\right)^{D} \, {\rm
e}^{-(r^2+s^2)/r_c^2}
\]

There is an alternative, more cumbersome but also more general,
method to handle spatially non-homogeneous systems, based on the
computation of
local correlation functions. It is instructive to verify that this
procedure leads to
results in agreement with the simple arguments above. In the language
of second quantization
the density is given by
\[
           {\hat\rho}(r) = \Psi^{\dag}(r)\,\Psi(r),
\]
while the local pressure can be expressed in terms of the trace of
the stress
tensor\footnote{With a momentum density
\(
  {\hat P}^i(x) = -\frac{i}{2}\hbar\left[\Psi^{\dag}\left(\partial_i
\Psi \right) -
     \left(\partial_i \Psi^{\dag}\right)\Psi
\right]
\),
and use of the equations of motion,
one finds that this expression for the stress tensor leads to a force
law of the
expected form:
\(
   \frac{\partial}{\partial t} {\hat P}^j + \partial_i {\hat T}^{ij}
= {\hat F}^j
\),
with ${\hat F}^j(r) = -\Psi^{\dag}(r)\,\Psi(r)\;\partial_j\Phi(r)$.
}
\begin{equation}
  {\hat T}^{ij}(r) = \frac{\hbar^2}{2m} \left\{
  \left(\partial_i\Psi^{\dag}(r)\right)\left(\partial_j\Psi(r)\right)
+
  \left(\partial_j\Psi^{\dag}(r)\right)\left(\partial_i\Psi(r)\right)
- \frac{1}{2}\delta_{ij}
  \nabla^2\left(\Psi^{\dag}(r)\Psi(r) \right)
  \right\},
\end{equation}
i.e.
\begin{equation}
   {\hat p}(r) = \frac{1}{D} \sum_{i=1}^{D} \;{\hat T}^{ii}(r).
\end{equation}
Here $\Psi(r)$ is the second quantized Schr{\"o}dinger field (assumed
to obey either
Bose or Fermi statistics),---thus ${\hat\rho}(r)$ and
${\hat T}_{ij}(r)$ are quantum
operators who's thermal average we want to compute.
This information is contained in the correlation function
\begin{eqnarray}
   &&G(r,s;\beta) =
   \mbox{\bf Tr}\left[ {\rm e}^{-\beta({\hat H}-\mu{\hat N})}\;
\Psi^{\dag}(r) \Psi(s) \right]/
   \mbox{\bf Tr}\left[ {\rm e}^{-\beta({\hat H}-\mu{\hat N})} \right]
=\nonumber\\&&
   \sum_{i} \frac{{\rm e}^{-\beta(\epsilon_i-\mu)}}{1\pm{\rm
e}^{\beta(\epsilon_i-\mu)}} \;
   \psi_i^{*}(r)\,\psi_i(s) =
   \sum_{\ell=1}^{\infty} (\pm 1)^{\ell+1} z^{\ell} \, \sum_i \;
   {\rm e}^{-\ell\beta\epsilon_i}\, \psi_i^{*}(r)\, \psi_i(s)
=\nonumber\\&&
   \sum_{\ell=1}^{\infty} (\pm1)^{\ell+1}\, z^{\ell}\;
   {\cal N}_{\ell}\,
   {\rm e}^{-\left({m\omega}/{2\hbar}\right)\left[
   (r^2+s^2)\coth(\ell\beta\hbar\omega) -
2rs/\sinh(\ell\beta\hbar\omega)
   \right] }.
\end{eqnarray}
where the sum over $i$ runs over all single particle energy levels,
with energies $\epsilon_i$
(we assume $\epsilon_{min} > \mu$) and normalized eigenfunctions
$\psi_i$.
${\cal N}_{\ell}=\left[m\omega/2\pi\hbar
\sinh(\ell\beta\hbar\omega)\right]^{D/2}$, and $r,\,s$
are $D$-dimensional vectors (their product means scalar product).
It follows that
\begin{eqnarray}
   &&\rho(r) = \langle \,{\hat\rho}(r) \,\rangle_{\beta} =
G(r,r,\beta) =
   \sum_{\ell=1}^{\infty} \,(\pm1)^{\ell+1}\, z^{\ell}\;
   {\cal N}_{\ell}\;
   {\rm e}^{-(r/r_{\ell})^2},
                                       \label{locdensity}
\end{eqnarray}
where $r_{\ell}^{-2} =
\left(\frac{m\omega}{\hbar}\right)
\tanh(\frac{1}{2}\ell\beta\hbar\omega)$.
Further
\begin{eqnarray}
   &&p(r) = \frac{\hbar^2}{2 m} \left\{\frac{2}{D}
   \left.\sum_{i=1}^D \partial_i^r\,\partial_i^s\, G(r,s;\beta)
\right|_{s\to r}
    - \frac{1}{2} \nabla^2 G(r,r;\beta) \right\} =\nonumber\\&&
   \frac{\hbar^2}{2m}\sum_{\ell=1}^{\infty} \,(\pm1)^{\ell+1}\,
z^{\ell}\;
   {\cal N}_{\ell}\,
   \left[ \frac{(2m\omega/\hbar)}{\sinh(\ell\beta\hbar\omega)} +
   \frac{D^2 r_{\ell}^2+2(1-D)r^2}{D r_{\ell}^4}  \right]
   \;
   {\rm e}^{-(r/r_{\ell})^2}.
                                        \label{locpressure}
\end{eqnarray}
In the limit $\omega\to 0^+$ equations (\ref{locpressure}) and
(\ref{locdensity}) turn into
the corresponding expressions in equation (\ref{pressure2}). The
parameter $\varepsilon$ which
determines the first correction to this limit expression for the
pressure is seen to be
\[
   \varepsilon = {\overline\ell}\beta\hbar\omega,\ \ \
   \mbox{with}\ \ \
   {\overline\ell}\beta p = z\frac{\partial}{\partial z} p = \rho.
\]
In conclusion, the harmonic oscillator regularization is a convenient
method to impose
particle confinement in a finite volume.
Taking into account the modifications due to the resulting spatial
nonhomogeneity
of the system all bulk thermodynamic quantities may in principle be
calculated,
at least perturbatively in a fugacity or density expansion.
Non-perturbatively the method may break down in regions with long
range correlations.


\newpage
\begin{thebibliography}{99}
\bibitem{Comtet}A.Comtet, Y.Georgelin and S.Ouvry, J.Phys. A\ {\bf
22}, 3917 (1989)
\bibitem{Leinaas}J.M.Leinaas and J.Myrheim, Nuovo Cimento B\ {\bf
37}, 1 (1977)
\bibitem{Wilczek}F.Wilczek, Phys.Rev.Lett.\ {\bf 49}, 957 (1982)
\bibitem{Arovas}D.P.Arovas, R.Schrieffer, F.Wilczek and A.Zee,
Nucl.Phys.\ {\bf B251}, 117 (1985)
\bibitem{Blum}T.Blum, C.R.Hagen and S.Ramaswamy, Phys.Rev.Lett.\ {\bf
64}, 709 (1990)
\bibitem{Sporre}M.Sporre, J.J.M.Verbaarschot and I.Zahed,
SUNY-NTG-91/47 (December 1991)
\bibitem{Sen}D.Sen, Phys.Rev.Lett.\ {\bf 68}, 2977 (1992)
\end{thebibliography}
\end{document}